\documentclass[aip,cha,preprint,amsmath,amssymb,groupedaddress,nofootinbib]{revtex4-1}
\usepackage[]{natbib}
\usepackage{amsmath, amssymb, amsthm,mathtools, url}
\usepackage[table]{xcolor}  % Allow for color text
\usepackage[normalem]{ulem} % Allow for strikeout text (\sout)
\usepackage{dcolumn}
\usepackage{setspace}
\usepackage{booktabs}
\newcolumntype{d}[1]{D{.}{.}{#1}}
\usepackage[utf8]{inputenc}
\usepackage{indentfirst}
\usepackage{graphicx}
\usepackage{float}

\graphicspath{ {./images/} }
\newcommand\scalemath[2]{\scalebox{#1}{\mbox{\ensuremath{\displaystyle #2}}}}

\begin{document}

\title{Modeling the Public Health Impact of E-Cigarettes on Adolescents and Adults}
\author{Lucia M.~Wagner}
\affiliation{Department of Mathematics, Statistics, and Computer Science\\
St.~Olaf College, Northfield, Minnesota, USA}

%\author{Gabby Digan}
%\affiliation{Department of Statistics \\ University of Illinois at Urbana-Champaign}

%\author{Ruiyi Wang}
%\affiliation{Department of Statistics \\ University of Chicago}

%\author{Elizabeth Wei}
%\affiliation{Department of Engineering Physics \\ University of Illinois at Urbana-Champaign}

\author{Sara M.~Clifton}
\email{clifto2@stolaf.edu}
\affiliation{Department of Mathematics, Statistics, and Computer Science\\
St.~Olaf College, Northfield, Minnesota, USA}

\date{\today}

\begin{abstract} 
Since the introduction of electronic cigarettes to the United States market in 2007, vaping prevalence has surged in both adult and adolescent populations. E-cigarettes are advertised as a safer alternative to traditional cigarettes and as a method of smoking cessation, but the U.S.~government and health professionals are concerned that e-cigarettes attract young non-smokers. Here, we develop and analyze a dynamical systems model of competition between traditional and electronic cigarettes for users. With this model, we predict the change in smoking prevalence due to the introduction of vaping, and we determine the conditions under which e-cigarettes present a net public health benefit or harm to society. 
\end{abstract} 


\maketitle

\spacing{1.25}
\section{Introduction}
Tobacco products, sources of the stimulant drug nicotine, are a serious detriment to health, causing damage to nearly all bodily organs \cite{yanbaeva2007systemic}. Inhaling tobacco smoke is a direct causative agent of disease --- including cancers, cardiac and pulmonary diseases, and diabetes --- and is known to increase risk for tuberculosis and immune system dysfunction \cite{us2014health}. Smoking persists as a leading preventable cause of premature death both in the U.S.~and globally \cite{us2014health}. Cigarette smoking alone is responsible for approximately one in five preventable deaths in the U.S.~annually \cite{us2014health}.

Through epidemiological studies, government policy and regulations, and advertising campaigns promoted by agencies such as the U.S.~Centers for Disease Control and Prevention (CDC), the American public has become increasingly aware of the negative repercussions of tobacco and, specifically, nicotine \cite{dube2009cigarette}. This cultural shift, driven largely by access to information about the dangers of tobacco, has ushered in a thriving market for cigarette smoking cessation products \cite{huang2019vaping}. In particular, electronic cigarettes (e-cigarettes) were introduced to the U.S.~market in 2007 as an alternative to traditional cigarettes \cite{perrine2019characteristics}. E-cigarettes are battery-powered devices that operate by emitting doses of vaporized nicotine that is inhaled by users. Often, the devices are discrete and the vapor is flavored \cite{perrine2019characteristics}. Because e-cigarettes have only recently been introduced to a mass market, the long-term effects on human health are largely unknown, but the general consensus by scientists and physicians is that smoking is riskier for one’s health than vaping \cite{goniewicz2018comparison}.

E-cigarette brands such as JUUL and SMOK advertise their products as a safer alternative to traditional cigarettes and as a means of smoking cessation \cite{huang2019vaping}. Although e-cigarettes may contain fewer hazardous chemicals than traditional cigarettes, e-cigarettes emit substantial quantities of nicotine \cite{goniewicz2018comparison} and emit known carcinogens \cite{hess2017cigarettes}. Nicotine is also highly addictive to both adult and adolescent users, and adolescents in particular are undergoing sensitive nervous system and hormone development \cite{yuan2015nicotine}. Nicotine negatively affects one’s brain activity, stimulating an unnatural secretion of chemical signals involved in the brain’s reward and pleasure systems \cite{benowitz2010nicotine}. Over time, this excessive secretion of chemicals leads to desensitization, tolerance of nicotine, and ultimately dependence on the drug \cite{benowitz2010nicotine}.

The vast range of vapor flavors, such as bubblegum or birthday cake, novelty of e-cigarettes, and discreteness of the devices disproportionately attract young nonsmokers \cite{soneji2019} Adolescence is a sensitive developmental stage, and this age group is particularly vulnerable to trying e-cigarettes; adolescents have low impulse control and are especially susceptible to social pressure \cite{yuan2015nicotine}. Further, a majority of adolescent e-cigarette users are unaware that vaporizers contain nicotine; a 2015 study reported that when asked the question \emph{``The last time you used an electronic vaporizer such as an e-cigarette, what was in the mist inhaled?''} nearly two-thirds of adolescents believed it was \emph{``just flavoring''} \cite{miech2017kids}.

Previous studies on traditional cigarette and e-cigarette consumption investigate prevalence in a descriptive or statistical manner \cite{hammond2019prevalence, steigerwald2018smoking, hammond2020changes}. Neither approach aims to predict future consumption, a result that would aid in governmental regulation of nicotine-emitting products. The mechanistic model presented here captures both the intrinsic and social utilities of smoking and vaping, with the goal of predicting future prevalence within the U.S. With this model, we investigate a counterfactual case in which vaping was never introduced to the U.S.~market, and we quantitatively analyze the net public health impact of e-cigarettes on society.

%%%%%%%%%%%%%%%%%%%%%%%%%%%%%%%%%%%%%%%%%%%%%%%%%%%%%%%%%%%%%%%%%%%%%%%%%%%%%%%%%%%%
\section{Methods}
\subsection{Model}
\begin{figure}[ht]
    \centering
\includegraphics[width=.7\textwidth]{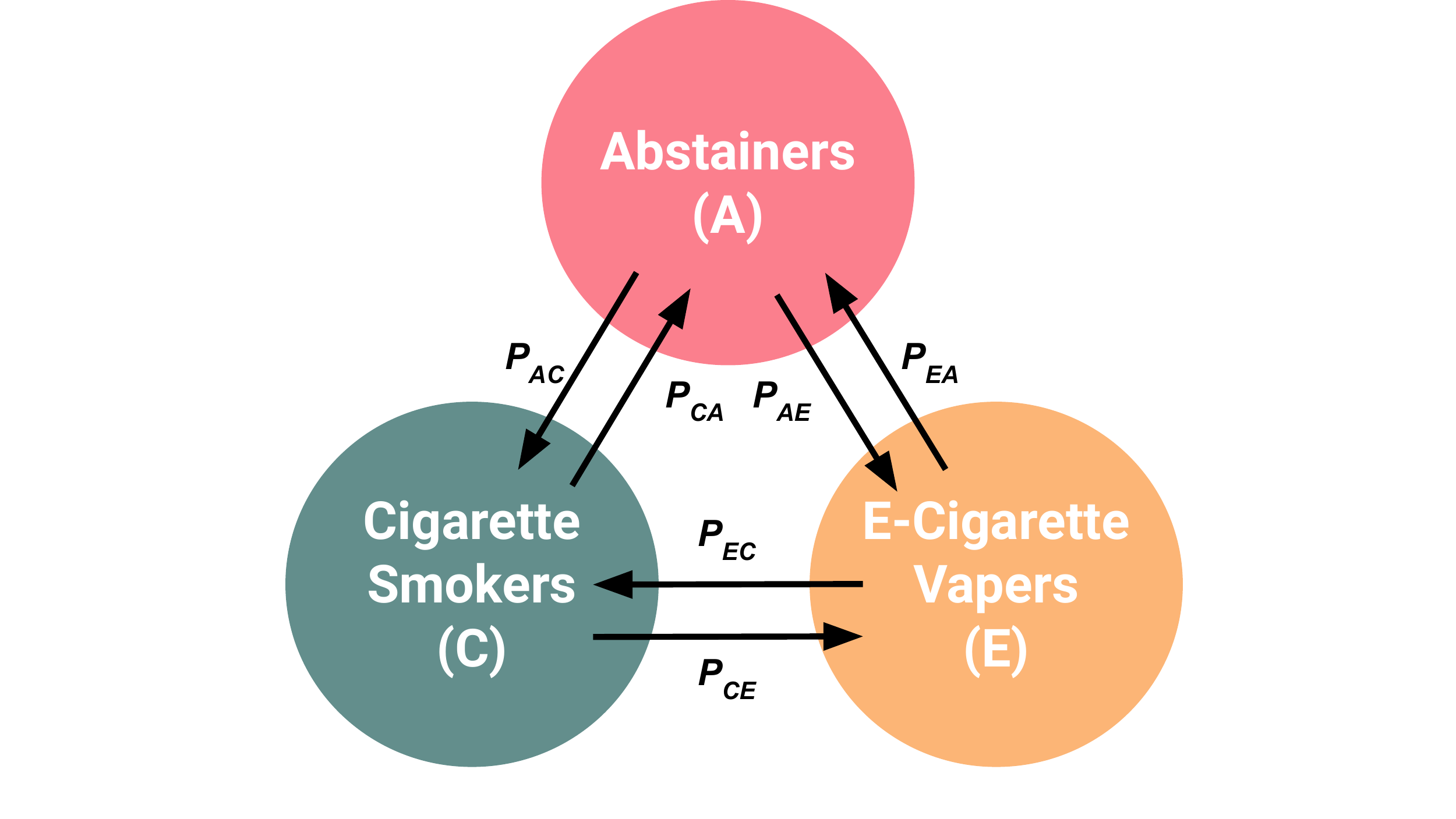}
    \caption{Compartmental diagram of model system (\ref{eq:dCdt})-(\ref{eq:dEdt}). Individuals transition among three discrete groups: those who abstain from tobacco products, those who primarily smoke cigarettes, and those who primarily vape e-cigarettes. Transitions are initiated based upon the intrinsic utilities and social utilities of the groups. The transition rates are modeled by (\ref{eq:Pyx}).}
    \label{fig:ACE}
\end{figure}

Because smoking and vaping are highly influenced by social cues, we use social group competition as the basis for our model \cite{abrams2011dynamics, wagener2016examining}. Social group competition models are utilized to study factions in society competing against each other for members, including religious groups \cite{abrams2011dynamics}, language speakers \cite{abrams2003modelling}, and even left- versus right-handedness \cite{abrams2012model}. A previous study by Lang et al.~used social group competition to understand smoking prevalence dynamics among adult populations around the world \cite{lang2015influence}. Here, we modify the Lang et al.~model to accommodate three discrete groups competing for members: individuals who primarily smoke traditional cigarettes, primarily vape e-cigarettes, and primarily abstain from those tobacco products (Figure \ref{fig:ACE}).

%%%%%%%%%%%%%%%%%%%%%%%%%%%%%%%%%%%%%%%%%%%%%%%%%%%%%%%%%%%%%%
%%%%%%%%%%%%%%%%%%%%%%%%%%%%%%%%%%%%%%%%%%%%%%%%%%%%%%%%%%%%%%

Our model tracks the transitions that occur among these three groups due to changing product utilities. The change in smoking prevalence $(C)$ and vaping prevalence $(E)$ over time is governed by
\begin{align}
\frac{\mathrm{d}C}{\mathrm{d}t} &= \underbrace{AP_{AC}(C; u_C)}_{\text{start smoking}}
+\underbrace{EP_{EC}(C; u_C)}_{\text{vaping to smoking}}- \underbrace{C P_{CA}(A; u_A)}_{\text{quit smoking}} - \underbrace{C P_{CE}(E; u_E)}_{\text{smoking to vaping}}
\label{eq:dCdt}\\
\frac{\mathrm{d}E}{\mathrm{d}t} &= \underbrace{AP_{AE}(E; u_E)}_{\text{start vaping}}
+ \underbrace{CP_{CE}(E; u_E)}_{\text{smoking to vaping}} - \underbrace{E P_{EA}(A; u_A)}_{\text{quit vaping}} - \underbrace{E P_{EC}(C; u_C)}_{\text{vaping to smoking}},
\label{eq:dEdt}
\end{align}
where $A = 1-C-E$ is the abstaining prevalence, $t$ is the time in years, $P_{YX}$ is the probability per unit time of transition from group $Y$ to group $X$, and $u_X$ is the intrinsic utility of group $X$.

We assume the probability of switching groups depends on both the \emph{intrinsic utility} and the \emph{social utility} of belonging to the group itself. The intrinsic utility includes all costs and benefits of belonging to a group that do not directly depend on the size of the group (e.g., the physiological response, addiction, monetary cost, and understanding of health risks). The social utility includes all costs and benefits of belonging to a group that depend on the popularity of the group (e.g., the sense of belonging, social protection, and conformity). Following Lang et al.~\cite{lang2015influence}, the probability per unit time of transitioning from group $Y$ to group $X$ is modeled by
\begin{align}
P_{YX}(x; u_x) = b \, x^{\alpha} \, u_x,
\label{eq:Pyx}
\end{align}
where $x$ is the fraction of the population in group $X$, $0\le u_x \le 1$ is the intrinsic utility of group $X$, $\alpha$ is a measure of societal conformity, and $b$ sets the timescale of transitions. 

Because knowledge about the dangers of smoking have increased over time, the intrinsic utility of smoking is a decreasing function of time. Lang et al.~estimate the intrinsic utility of smoking using the cumulative number of scientific articles published on the link between smoking and cancer, discounted by the saturation of the public's attention \cite{lang2015influence}. The resulting intrinsic utility function is approximately sigmoidal, so we use the following sigmoid function to simplify their data-driven approach: 
\begin{align}
u_x(t) = u_x^{0}+\frac{u_x^{\infty}-u_x^{0}}{1+ e^{-\lambda (t-T_x)}},
\label{eq:ux}
\end{align}
where $\lambda$ represents the rate at which people respond to new knowledge, $u_x^0$ represents the initial intrinsic utility of group $X$, $u_x^{\infty}$ represents the final intrinsic utility of group $X$, and $T_x$ is the inflection point of new knowledge (i.e., the year in which intrinsic utility changes most rapidly). Refer to the Supplemental Information for a visualization of Equation \ref{eq:ux}.

See Table \ref{tab:param} for a summary of model variables, parameters, and their fitted values. Although we are primarily interested in the transient dynamics of the model, we perform a steady-state analysis for the sake of thoroughness (see Supplemental Information).

%%%%%%%%% Parameter table %%%%%%%%
	\begin{table}[ht]
	\caption{Description of model variables and parameters in system (\ref{eq:dCdt})-(\ref{eq:dEdt}). Fitted values for adult prevalence indicated with (a), and fitted values for youth prevalence indicated with (y).}
\footnotesize
\renewcommand{\arraystretch}{1.1}
	\begin{tabular}{| c  p{9.1cm}  p{2.9cm} c |}  \hline 
	{\bf } & {\bf Meaning} & {\bf Value} & {\bf Sources} \\  \hline 
	$C$ & proportion of population that primarily smokes traditional cigarettes & Variable & -- \\ 
	$E$ & proportion of population that primarily vapes e-cigarettes & Variable & -- \\
	$A$ & proportion of population that primarily abstains from nicotine products & Variable & -- \\
    $t$ & time in years & Variable & -- \\
    $\displaystyle u_{A},u_{C},u_{E}$ & intrinsic utilities of abstaining, smoking, and vaping, respectively (unitless quantities such that $u_A+u_C+u_E=1$) & Variable & -- \\
    $\alpha$ & degree of societal conformity (U.S.) & 0.963 & \cite{lang2015influence} \\
    $b$ & timescale of transitions & 1.0 & \cite{lang2015influence} \\
    $\displaystyle u_{C}^0$ & initial ($t\to-\infty$) intrinsic utility of smoking & 0.52 (a) 0.55 (y) & \cite{lang2015influence}, this study \\
    $\displaystyle u_C^{\infty}$ & final ($t\to\infty$) intrinsic utility of smoking & 0.27 (a) 0.16 (y) & this study \\
    $\displaystyle u_C^{\infty*}$ & final ($t\to\infty$) intrinsic utility of smoking had e-cigarettes never been introduced (counterfactual) & 0.48 (a) 0.46 (y) &  \cite{lang2015influence}, this study \\
    $\displaystyle u_{E}^0$ & initial ($t\to-\infty$) intrinsic utility of vaping & 0.41 (a) 0.56 (y) & this study \\
    $\displaystyle u_E^{\infty}$ & final ($t\to\infty$) intrinsic utility of vaping & 0.34 (a) 0.01 (y) & this study \\
    $\lambda$ & response rate to new information (per year) & 0.10 (a) 0.83 (y) & this study \\
    $T_C$ & year of maximum change in cigarette utility & 1964 (a), 1995 (y) & \cite{lang2015influence,us2014health} \\
    $T_E$ & year of maximum change in e-cigarette utility & 2025 & guess$^{\dagger}$ \\
	\hline
	\end{tabular}
	\label{tab:param}
	$^{\dagger}$Sensitivity analysis for this parameter in Results
	\end{table}
%%%%%%%%% Parameter table %%%%%%%%

\subsection{Prevalence Data}
Before 1992, the CDC defined a current cigarette smoker as an individual who reported having smoked at least 100 cigarettes in their lifetime and who smokes ``now" \cite{centers1992cigarette}. From 1992 onward, the CDC defines a current smoker as an individual who has smoked at least 100 cigarettes in their lifetime, and who currently smokes cigarettes either ``everyday" or ``some days" \cite{dube2009cigarette}. The CDC has consistently defined a current user of e-cigarettes as one who vapes ``on one or more of the past 30 days" \cite{wang2019tobacco,tobaccotrendsbrief}.

To quantify adult smoking prevalence over time, we use data from the CDC when available \cite{tobaccotrendsbrief}. When yearly adult smoking prevalence was not available from the CDC, we use estimates from Lang et al.~based on total cigarette consumption \cite{lang2015influence}. Data for adult smoking span the years 1920 to 2019. Data for youth smoking\footnote{For our purposes, ``youths" or ``adolescents" are defined as high school students.} spans from 1991 to 2019, obtained from the American Lung Association's analysis of CDC data \cite{tobaccotrendsbrief}.

\begin{figure}[ht]
    \centering
\includegraphics[width=0.85\textwidth]{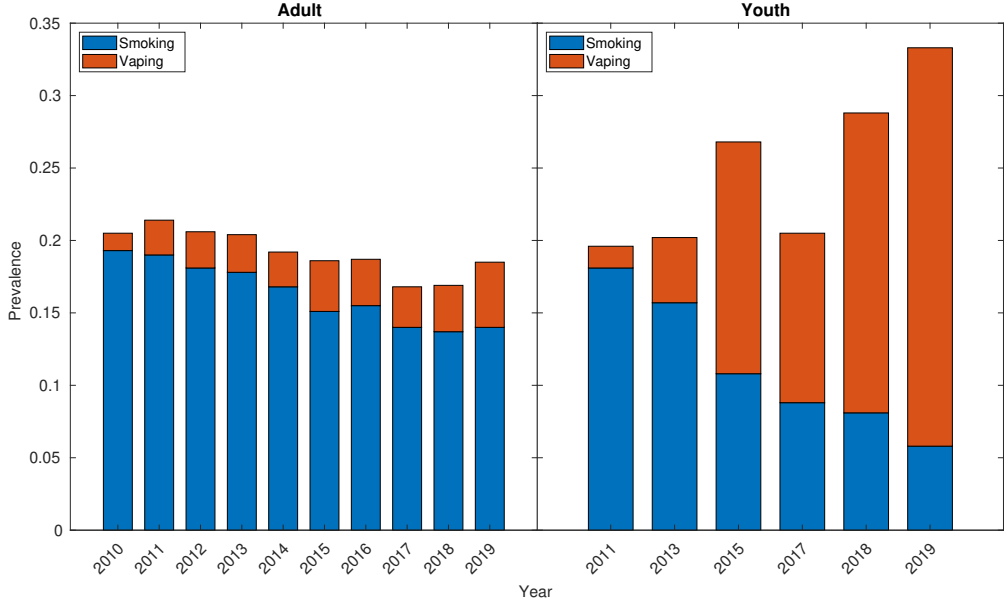}
    \caption{Relative prevalence for smoking and vaping for adults and youths in U.S. society. There is a steady rise in the prevalence of e-cigarette consumption by adults in recent years, which counteracts the decreasing trend of traditional cigarette consumption. There is a rapid increase in the prevalence of e-cigarette consumption by youths in recent years, leading to an overall increase in tobacco use despite the rapid decline in youth smoking.}
    \label{fig:adultyouthprev}
\end{figure}

Data for adult and youth vaping span the years 2010 to 2019 and 2011 to 2019, respectively, also from the American Lung Association's analysis of CDC data \cite{tobaccotrendsbrief}. See Figure \ref{fig:adultyouthprev} for a visualization of total smoking and vaping prevalence after the introduction of e-cigarettes.

%%%%%%%%%%%%%%%%%%%%%%%%%%%%%%%%%%%%%%%%%%%%%%%%%%%%%%%%%%%%%%
%%%%%%%%%%%%%%%%%%%%%%%%%%%%%%%%%%%%%%%%%%%%%%%%%%%%%%%%%%%%%%

\subsection{Fixed Parameter Values}
When fitting our model, we assumed the year of maximum change in cigarette utility ($T_C$) for adult users was 1964, based on the inflection point seen in the intrinsic utility function in Lang et al.~\cite{lang2015influence}. This year is plausible because in 1962, the Royal College of Physicians in England directly communicated that cigarettes cause lung cancer \cite{us2014health}. Shortly thereafter, in 1964, the U.S.~Surgeon General reached a similar conclusion in a widely publicised report \cite{us2014health}. The model fit to adult prevalence data is not especially sensitive to the precise year selected, as the model looks nearly indistinguishable for all $T_C \in [1952,1978]$. 

For youths, we assumed the year of maximum change in cigarette utility ($T_C$) was 1995, when the U.S.~Food and Drug Administration declared nicotine a drug, ushering in a new wave of media coverage on the dangers of tobacco \cite{us2014health}. The model fit is more sensitive to this choice, likely because the data are more sparse for youths. The model fit is poor outside the range $T_C \in [1991,1996]$.

The degree of societal conformity ($\alpha$) is different for every society, with more individualistic societies having a smaller $\alpha$ and more collectivist societies having a larger $\alpha$. Lang et al.~estimate (and rigorously confirm) that $\alpha = 0.963$ for the U.S.~\cite{lang2015influence}. Likewise, the timescale parameter ($b$) is estimated by Lang et al.~to be $b=1.0$ per year \cite{lang2015influence}. 

The year of maximum change in e-cigarette utility ($T_E$) is unknown and cannot be fitted because the peak of vaping prevalence has not yet occurred. For the purposes of model fitting, we set $T_E = 2025$ for both adult and adolescent populations, and we perform a sensitivity analysis for this parameter.

\subsection{Parameter Estimation}
The remaining model parameters are fitted to smoking and vaping prevalence data; we assume the remaining parameters may differ between adult and youth populations due to both physiological and social differences. Best fit parameters were estimated by minimizing the error between the model and the data using the Nelder-Mead algorithm \cite{nelder1965simplex}. To reduce the risk of overfitting, we fit the model to each dataset in two steps. 

First, we fit to smoking data from the beginning of CDC data collection until the year when vaping was introduced. The parameters fit comprise the initial utility of smoking ($u_C^{0}$) and final utility of smoking ($u_C^{\infty *}$) if vaping were never introduced, the rate at which new knowledge is created and absorbed ($\lambda$), and the initial prevalence of smokers ($C(0)$).  

Next, we simultaneously fit to smoking and vaping data from the first year with vaping data through the last year of the dataset. The parameters fit comprise the final utility of smoking ($u_C^{\infty}$), the initial utility of vaping ($u_E^{0}$), the final utility of vaping ($u_E^{\infty}$), and the initial prevalence of vaping ($E(0)$). 

%%%%%%%%%%%%%%%%%%%%%%%%%%%%%%%%%%%%%%%%%%%%%%%%%%%%%%%%%%%%%%%%%%%%%%%%%%%%%%%%%%%%
\section{Results}
\subsection{Model Fits and Projections}
\begin{figure}[ht]
    \centering
    \includegraphics[width=.75\textwidth]{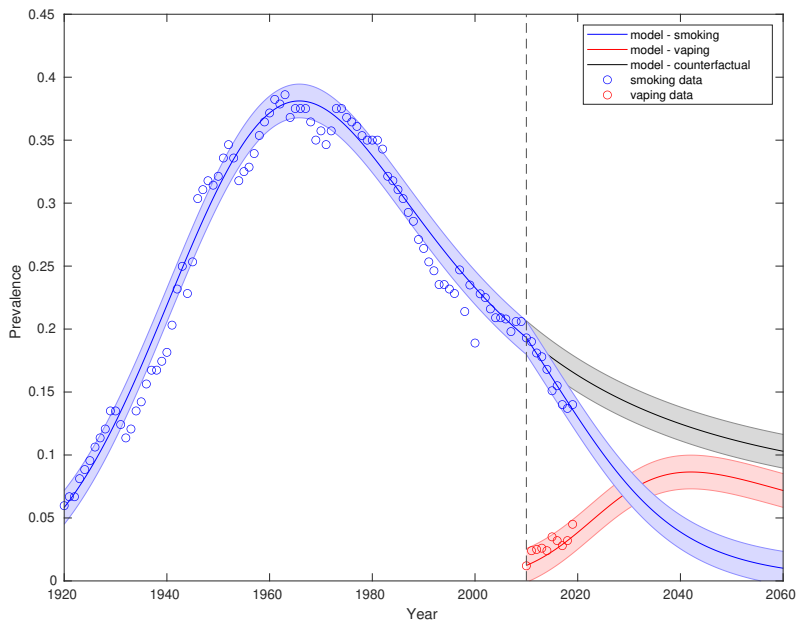}
    \caption{Model projections and data for U.S. adult smoking and vaping. The counterfactual, which assumes that vaping was never introduced, indicates that e-cigarettes had a clear impact on the consumption of traditional cigarettes. The dashed line marks the year (2010) that e-cigarette data was first collected by the CDC. The shaded envelope is $\pm$ average residual.} 
    \label{fig:adult}
\end{figure}

The model prediction agrees with adult smoking prevalence data that the peak prevalence of about 38\% occurred around 1965 (Figure \ref{fig:adult}). In the counterfactual scenario in which vaping had never been introduced, the model predicts that adult smoking prevalence would be higher than with vaping available as an alternative (Figure \ref{fig:adult}). 
All else held constant, the model projection estimates that adult smoking will reach a prevalence less than 1\% by the early 2060s.
The model projects that the peak of e-cigarette usage will occur around 2040 with a prevalence a little under 10\%.

%%%%%%%%%%%%%%%%%%%%%%%%%%%%%%%%%%%%%%%%%%%%%%%%%%%%%%%%%%%%%%%%%%
%%%%%%%%%%%%%%%%%%%%%%%%%%%%%%%%%%%%%%%%%%%%%%%%%%%%%%%%%%%%%%%%%%

From 2010, the year e-cigarette prevalence data for adults started being collected by the CDC, to 2030, the average prevalence per year of adult smokers diverted by vaping is estimated to be 0.35\%. To put that prevalence into perspective based on the CDC's estimate of U.S. adults who currently smoke cigarettes \cite{cdc2019current}, that is nearly 120,000 individuals per year that transition out of the cigarette group to either to the abstainer group or, more likely, to the e-cigarette group.

\begin{figure}[ht]
    \centering
    \includegraphics[width=.75\textwidth]{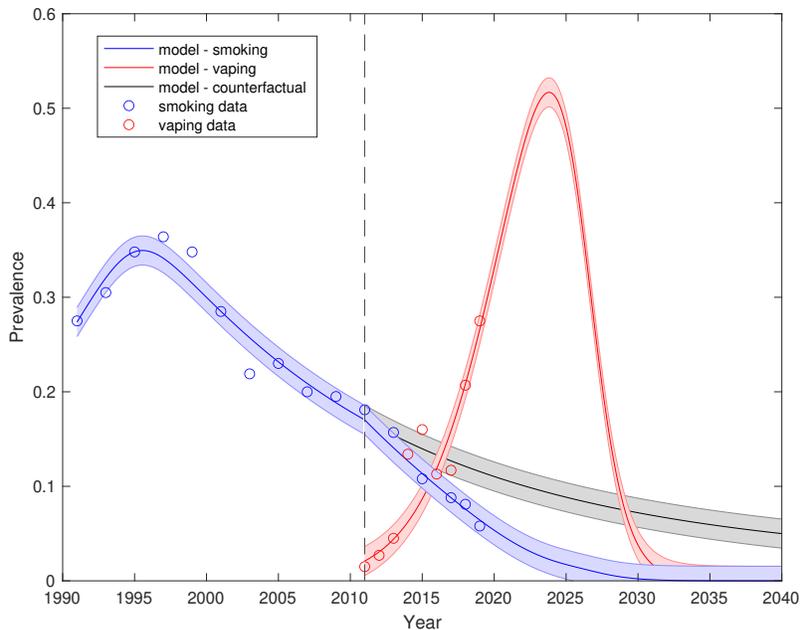}
    \caption{Model projections and data for U.S. adolescent smoking and vaping. The counterfactual, which assumes that vaping was never introduced, indicates that e-cigarettes had a clear impact on the consumption of traditional cigarettes for adolescents. The dashed line marks the year (2011) that e-cigarette data was first collected by the CDC for young users. The shaded envelope is $\pm$ average residual.}
    \label{fig:youth}
\end{figure}

The model predicts that the peak of traditional cigarette consumption among adolescents occurred around 1996 with a prevalence of about 35\%; this estimate might be slightly early and low (Figure \ref{fig:youth}). The model suggests that the introduction of e-cigarettes also decreased traditional cigarette prevalence for youths. Adolescent smoking in the U.S. is projected to reach a prevalence less than 1\% in the mid-2020s, whereas adolescent vaping is predicted to reach this prevalence in the early 2030s. The peak e-cigarette usage is projected to occur around 2025 with a prevalence of about 52\%.

From 2011, the year e-cigarette prevalence data for high school students started being collected by the CDC, to 2030, the average prevalence per year of adolescent smokers diverted by vaping is estimated to be 0.39\%. Based on the current number of high school students in the U.S. \cite{statshighschool} and the the CDC's estimate of U.S. high schoolers who currently smoke cigarettes \cite{tobaccotrendsbrief}, that is nearly 58,000 individuals per year that are projected to transition out of the cigarette group on account of vaping. 

%%%%%%%%%%%%%%%%%%%%%%%%%%%%%%%%%%%%%%%%%%%%%%%%%%%%%%%%%%%%%%%%%%
%%%%%%%%%%%%%%%%%%%%%%%%%%%%%%%%%%%%%%%%%%%%%%%%%%%%%%%%%%%%%%%%%%

\subsection{Public Health Costs and Benefits}
While the model predicts that the introduction of vaping decreased smoking prevalence for both adult and adolescent populations, that does not necessarily mean that vaping has a net positive impact on public health. If we suppose that the model projections resemble reality, we can estimate the net public health benefit or harm caused by the introduction of e-cigarettes to the U.S.~market. The net public health benefit or harm depends on two factors: (1) how many e-cigarette users were former smokers versus former abstainers, and (2) the relative health risk of smoking versus vaping. Unfortunately, the long-term effects of vaping are largely unknown due to the recent introduction of e-cigarette devices. However, the current consensus hypothesis among scientists and physicians is that vaping is less harmful than smoking \cite{goniewicz2018comparison}. 

We define the net public health cost/benefit of nicotine products at any given time $t$ since the introduction of e-cigarettes as the average change in smoking prevalence due to vaping plus the average prevalence of vaping, weighted by the relative health risk of smoking versus vaping. For instance, suppose that for every smoker diverted to vaping another abstainer is also diverted to vaping. If smoking is more than two times riskier than vaping, then vaping is a net public health gain. Otherwise vaping is a net public health loss.

\begin{figure}[ht]
    \centering
    \includegraphics[width=\textwidth]{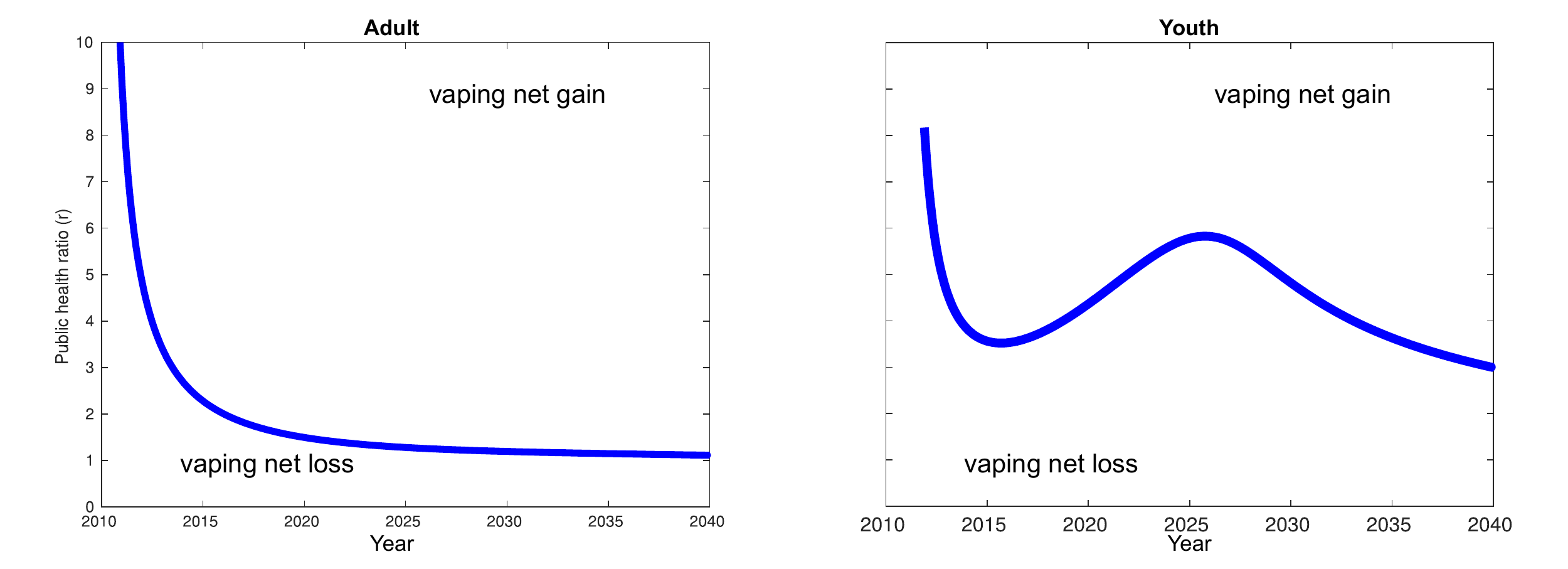}
    \caption{Public health impact of e-cigarettes for adults and youths in the U.S. The ratio $r$ measures how many times riskier smoking is relative to vaping. This value is currently unknown. The blue curve ($\Delta c = 0$) is the ratio for which vaping has a net neutral impact on society based on model predictions. For example, in the year 2030, if the risk of smoking is greater than 1.2 times worse than vaping, e-cigarettes present a net benefit for public health in adult populations. For youths in the year 2030, e-cigarettes present a net benefit if smoking is more than 4.8 times riskier.}
    \label{fig:publichealthratio}
\end{figure}

Mathematically, we define the public health cost of nicotine products to be 
\begin{align}
c(t) = 0\cdot \bar{A}(t) + r \cdot \bar{C}(t) + 1\cdot \bar{E}(t), 
\label{eq:cost}
\end{align}
where the risk due to abstaining is 0, the risk due to vaping is arbitrarily set to 1, the risk due to smoking is $r$ (a ratio of smoking to vaping risk), and $\bar{A}(t),\bar{C}(t),\bar{E}(t)$ are the time-averaged prevalence of abstaining, smoking, and vaping since the introduction of e-cigarettes\footnote{For simplicity, this formulation ignores the average consumption of tobacco products per user. Higher consumption of either product per user is likely to cause more harm, for which we do not account.}. In a counterfactual scenario where vaping was never introduced, the public health cost is 
\begin{align}
\tilde{c}(t) = 0\cdot \tilde{\bar{A}}(t) + r \cdot \tilde{\bar{C}}(t), 
\label{eq:costcf}
\end{align}
where $\tilde{\bar{A}}(t)$ and $\tilde{\bar{C}}(t)$ are the model projections for abstaining and smoking prevalence (time-averaged since the introduction of vaping) in the counterfactual scenario in which vaping is never introduced ($\tilde{\bar{E}}(t) \equiv 0$). We are interested in the cost difference between the counterfactual and actual scenarios:
\begin{align}
\Delta c(t) = r \cdot \tilde{\bar{C}}(t) - (r \cdot \bar{C}(t) + 1\cdot \bar{E}(t)) = r \big(\tilde{\bar{C}}(t) - \bar{C}(t)\big) - \bar{E}(t) 
\label{eq:deltac}
\end{align}
Here, $\Delta c > 0$ implies the introduction of vaping is on net harmful to public health, and $\Delta c < 0$ implies that the introduction of vaping is on net beneficial to public health. Because the relative risk of smoking to vaping ($r$) is currently unknown, we plot the regions in the $t - r$ plane where $\Delta c > 0$ (net harm of vaping) and $\Delta c < 0$ (net benefit of vaping); see Figure \ref{fig:publichealthratio}.

%%%%%%%%%%%%%%%%%%%%%%%%%%%%%%%%%%%%%%%%%%%%%%%%%%%%%%%%%%%%%%%%%%
%%%%%%%%%%%%%%%%%%%%%%%%%%%%%%%%%%%%%%%%%%%%%%%%%%%%%%%%%%%%%%%%%%

The public health cost function suggests that in the year 2030, if the health risk of smoking is at least 1.2 times worse than vaping, e-cigarettes present a net benefit for public health in adult populations. As for adolescent populations in the year 2030, if the health risk of smoking is at least 4.8 times worse than vaping, e-cigarettes present a net benefit for public health in adolescent populations. This ratio generally decreases over time, as the vaping fad comes to an end relatively quickly while the gap between smoking prevalence in the actual and counterfactual scenarios slowly closes.

\subsection{Sensitivity Analysis}
Because physicians and epidemiologists have not yet concluded the long-term health risks of vaping, the year in which the intrinsic utility of vaping will change most rapidly ($T_E$) has likely not occurred yet. Therefore the vaping inflection year $T_E$ is unknown and cannot be estimated using available data, so we perform a sensitivity analysis on the impact of $T_E$ on key findings (Figure \ref{fig:ecig}). See Supplemental Information for details.  

%%%%%%%%%%%%%%%%%%%%%%%%%%%%%%%%%%%%%%%%%%%%%%%%%%%%%%%%%%%%%
\begin{figure}[ht]
    \centering
    \includegraphics[width=\textwidth]{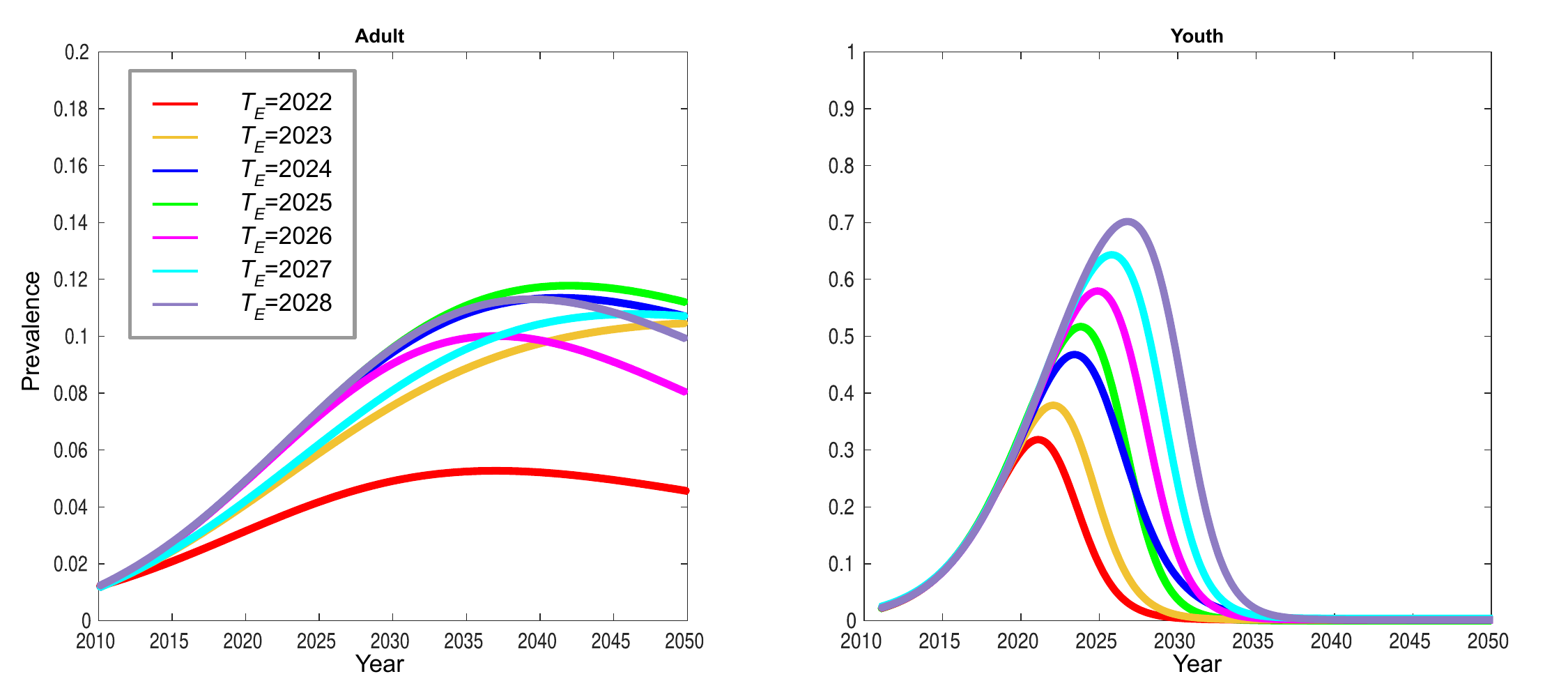}
    \caption{The model projections for vaping depending on the e-cigarette consumption inflection year ($T_E$). Over the range of plausible $T_E$ values, the projections all estimate that the peak prevalence of e-cigarette consumption by adolescents will occur sooner than for adults. However, the peak prevalence of vaping is predicted to be higher for adolescents than adults.}
    \label{fig:ecig}
\end{figure}  
%%%%%%%%%%%%%%%%%%%%%%%%%%%%%%%%%%%%%%%%%%%%%%%%%%%%%%%%%%%%%

\begin{figure}[htb]
    \centering
    \includegraphics[width=.9\textwidth]{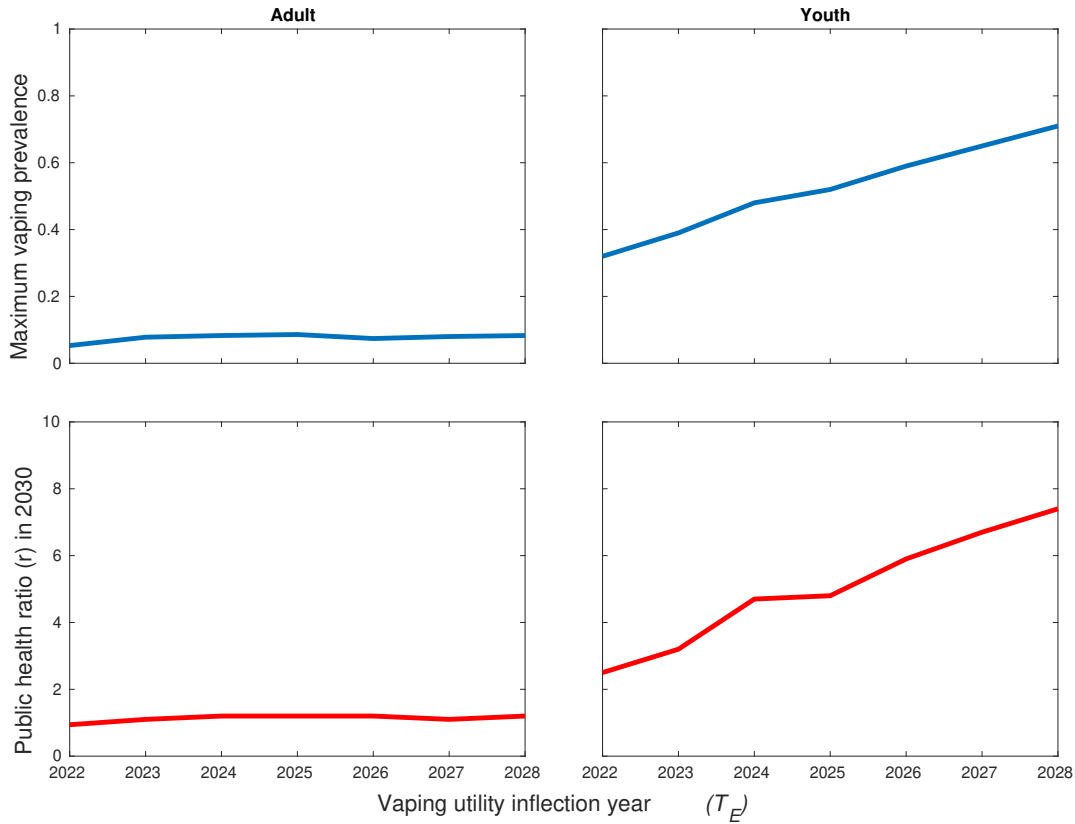}
    \caption{Sensitivity analysis of e-cigarette utility inflection point year ($T_E$) on peak vaping prevalence (top row) and net neutral public health ratio (bottom row). Predictions for youths are more sensitive to the vaping inflection year than predictions for adults.}
    \label{fig:sense}
\end{figure}

As expected, the peak prevalence of vaping for both adults and youths is sensitive to the year in which the intrinsic utility of vaping changes most rapidly (see Figure \ref{fig:sense}). The inflection point year precedes the year of peak prevalence, where the stability of the fixed point changes (see Supplemental Information). While the peak vaping prevalence for adolescents increases steadily from 30\% to 70\% as $T_E$ increases, the peak prevalence for adults remains in the range 5-12\% as $T_E$ increases, but without a discernible pattern. 

In Figure \ref{fig:sense}, we also estimate the net neutral public health ratio $r$ ($\Delta c = 0$) for adolescents and adults in the year 2030 over a range of plausible vaping inflection years $T_E$. As expected, this ratio $r$ is highly correlated with the peak vaping prevalence. Both metrics show that youth vaping predictions are more sensitive to $T_E$ than adult vaping predictions, suggesting that interventions that could rapidly decrease the intrinsic utility of vaping for youths (e.g., anti-vaping campaigns, regulations, etc.) should be implemented sooner rather than later. 

%%%%%%%%%%%%%%%%%%%%%%%%%%%%%%%%%%%%%%%%%%
%%%%%%%%%%%%%%%%%%%%%%%%%%%%%%%%%%%%%%%%%%

 %We would expect that as the vaping inflection year increases, the peak vaping prevalence would also increase, as the system has more time to approach the vaping-only fixed point. While we observe this for youth vaping, the peak prevalence for adults is relatively steady (Figures \ref{fig:ecig} and \ref{fig:sense}). This is likely because more data are available for adult prevalence.

%%%%%%%%%%%%%%%%%%%%%%%%%%%%%%%%%%%%%%%%%%%%%%%%%%%%%%%%%%%%%%%%%%%%%%%%%%%%%%%%%%%%
\section{Discussion}
The model projection estimates that the introduction of e-cigarettes into the U.S. market decreased traditional cigarette consumption among adults, confirming previous studies that adults use vaping products as a method of smoking cessation \cite{zhuang2016long}. Studies report that the vast majority of adults who use electronic delivery devices such as e-cigarettes do so in order to quit smoking \cite{tackett2015biochemically, warner2019cigarettes}. In 2015, 68\% of adult smokers wanted to quit and more than half had made a quit attempt in the past year, yet fewer than 10\% were successful \cite{tobaccotrendsbrief}. Reasons for this failure often include effects of withdrawal, sensory triggers, inadequate social support, and nicotine addiction \cite{benowitz2010nicotine}. With the introduction of e-cigarettes, adult smokers secured a new opportunity to acquire nicotine boosts while likely switching to a safer alternative. 

Similarly, smoking prevalence among adolescents decreased after the introduction of vaping. However, research suggests that the lure of e-cigarettes for adolescents is less spurred by pressure to adopt a safer lifestyle and more so by evolving social norms \cite{harrell2019vaping}. Based on the vaping projections illustrated in Figure \ref{fig:ecig}, our model estimates that the peak prevalence of e-cigarette consumption by adolescents will occur sooner than for adults, but at a higher prevalence. Further, according to the prevalence data reported by the CDC and represented by Figure \ref{fig:adultyouthprev}, nicotine-emitting products are more popular among adolescents than adults, a recent phenomena. These findings support the idea that adolescents are more so attracted to vaping based on current cultural attitudes towards the devices.

Other differences between adult and adolescent use of traditional and electironic cigarettes are evident in the fitted model parameters as well (Table \ref{tab:param}). We find there is little difference between the adult and youth initial intrinsic utility of smoking $(u_{C}^0$) and final intrinsic utility of smoking without e-cigarettes (counterfactual, $u_C^{\infty*}$), and these parameters are nearly identical to those found by Lang et al.~\cite{lang2015influence} using a different approach. There are, however, differences between adult and youth parameters once vaping is introduced to the U.S.~market. 

The final intrinsic utility of smoking after the introduction of e-cigarettes ($u_C^{\infty}$) is greater for adults than for adolescents. This finding is confirmed by the literature showing adolescents believe more strongly than adults that traditional cigarettes are harmful and addictive \cite{amrock2016perceptions}. 

The initial intrinsic utility of vaping ($u_{E}^0$) is estimated to be greater for adolescents than for adults. This is reasonable because adolescents are more likely than adults to develop nicotine addiction \cite{schramm2009adolescents} and possess less knowledge on the health risks of inhaling nicotine via vaporizers \cite{pepper2018adolescents}.  

The final intrinsic utility of vaping ($u_E^{\infty}$) for adults is substantially larger than that for adolescents. Because adolescents are more likely drawn to vaping based on current cultural trends and peer influence \cite{harrell2019vaping}, it is plausible that youths would perceive a low intrinsic value of vaping once the trend is over. On the other hand, adults are more likely to use vaping as a smoking cessation device, which holds intrinsic value long-term.

Our sensitivity analysis on the parameter $T_E$ may offer suggestions for future policy as well. For example, the rapid rise in youth vaping peak prevalence as the vaping inflection point increases suggests that government interventions are needed sooner rather than later. If government agencies delay serious action to decrease e-cigarette consumption, not only will the peak prevalence increase but there is a decreased chance that vaping will be a net benefit for public health in society. 

\subsection{Limitations}
Due to simplifying assumptions regarding collective human behavior used to create our model, predictions should be interpreted with caution. To begin, the model assumes people are aware of the current smoking prevalence nationwide and respond via social pressure. This is not a major limitation because assuming an all-to-all social network is not qualitatively different from assuming a more realistic social network in social group competition models \cite{abrams2011dynamics}. 

We also assume that adult and adolescent populations are independent and do not socially influence one another. While it may be reasonable to assume that adults and adolescents primarily respond to behaviors of others in their own age group \cite{knoll2017age}, the social influence on a child by a smoking parent may not be negligible \cite{jackson1997say}. 

Another simplifying assumption is that smoking and vaping are mutually exclusive. However, dual use of traditional and electronic cigarettes may not be negligible, especially for individuals using e-cigarettes as a smoking cessation device \cite{owusu2019patterns, wills2015risk}. A more sophisticated model would include dual use, but it would be more challenging to validate because the CDC does not consistently collect dual use data.

We also assume that the intrinsic utility function is monotonic in time, with the change over time depending primarily on awareness of health risks caused by nicotine delivery devices. In reality, the intrinsic utility of smoking and vaping may fluctuate over time due to government regulations, anti-nicotine campaigns, and major public health events. For instance, the U.S.~government banned flavored vapors in 2020 \cite{tanne2020fda}; this regulation (and the preceding bad publicity) caused the valuation of JUUL, which comprises 68\% of the U.S. e-cigarette market, to decrease from \$38 billion to less than \$5 billion between 2018 and 2020 \cite{maloney_2020}. Also, major public health events relevant to lung health such as e-cigarette/vaping-associated lung injury (EVALI) \cite{patel2020patient} and COVID-19 \cite{gaiha2020association} are not included in the model.

Limitations in data availability also encourage caution in interpreting model predictions. First, all prevalence data is based on sampling, which inherently involves uncertainty. Second, because the CDC only began collecting data on vaping in the early 2010s, vaping trends are not fully resolved and overfitting is a risk. Third, smoking prevalence data for adults comes from two different sources (historical consumption \cite{lang2015influence} and surveys \cite{tobaccotrendsbrief}); although the qualitative trends agree when the two datasets overlap, these differing source methodologies may introduce additional uncertainty. Finally, data for youth smoking is sparse; prevalence data only goes back to the mid-1990s with many missing years in between.

%%%%%%%%%%%%%%%%%%%%%%%%%%%%%%%%%%%%%%%%%%%%%%%%%%%%%%%%%%%%%%%%%%%%%%%%%%%%%%%%%%%%
\section{Conclusion}
Using a simple model of competition between traditional cigarettes and e-cigarettes for users, we predict the change in smoking prevalence due to the introduction of vaping in the U.S. Vaping products appear to decrease the prevalence of smoking among both adult and adolescent populations. Because the long-term health risks of vaping are currently unknown, the public health cost and/or benefit of e-cigarette is less clear. However, as suggested by our model, it is possible that the introduction of e-cigarettes will have positive effects for both adult and youth populations, depending on the relative health risk of smoking and vaping.

\section{Data Availability}
All data (Excel spreadsheets and MATLAB scripts) will be made available from the corresponding author by request.

\section{Acknowledgements}
The authors thank Gabby Digan, Ruiyi Wang, and Elizabeth Wei for contributions to early exploration of the model. Thanks are also due to the Illinois Geometry Lab and Mathways Grant No. DMS-1449269 (SMC) for research support.

\bibliographystyle{ieeetr}
\bibliography{bibliography}

\setcounter{section}{0}
\renewcommand{\thesection}{S\arabic{section}}
\setcounter{equation}{0}
\renewcommand{\theequation}{S\arabic{equation}}
\setcounter{figure}{0}
\renewcommand{\theequation}{S\arabic{figure}}
\setcounter{table}{0}
\renewcommand{\theequation}{S\arabic{table}}
\newpage
\section{Supplementary Information}
\subsection{Steady State Analysis}
In order to understand model behavior, we perform an equilibrium analysis of our differential equation system (\ref{eq:dCdt})-(\ref{eq:dEdt}). At steady state, each utility tends towards its final utility $u_x^{\infty}$, a constant; we suppress the infinity notation in this section for clarity. 

In the case $\alpha < 1$, the system is not analytically tractable except for a few special cases, and the equilibria cannot be written in closed form. However, stable coexistence of groups is expected to be possible \cite{abrams2011dynamics}, and we see that behavior in our model. Further analysis in this case is outside the scope of our paper. 

\subsubsection{Special Case: $\alpha = 1$}
Under the simplifying condition that $\alpha=1$ (slightly more collectivist society than the U.S.), the three equilibrium points are $(C^*,E^*) = (0,0), (0,1)$, and $(1,0)$. This indicates that one of the groups will dominate while the other two groups will ultimately cease to exist. 

As our differential equations are nonlinear, we utilize linear stability analysis to classify the equilibria. The Jacobian matrix is:
\[\scalemath{0.85}{{J =
\begin{bmatrix}
u_C - 2Cu_C - E u_E +(E+2C-1)(1-u_C-u_E) & -Cu_E+C(1-u_C-u_E)\\
-Eu_C+E(1-u_C-u_E)& u_E - 2Eu_E - C u_C +(C+2E-1)(1-u_C-u_E) \\
\end{bmatrix}}}
\]

\begin{figure}[th]
    \centering
    \includegraphics[width=.5\textwidth]{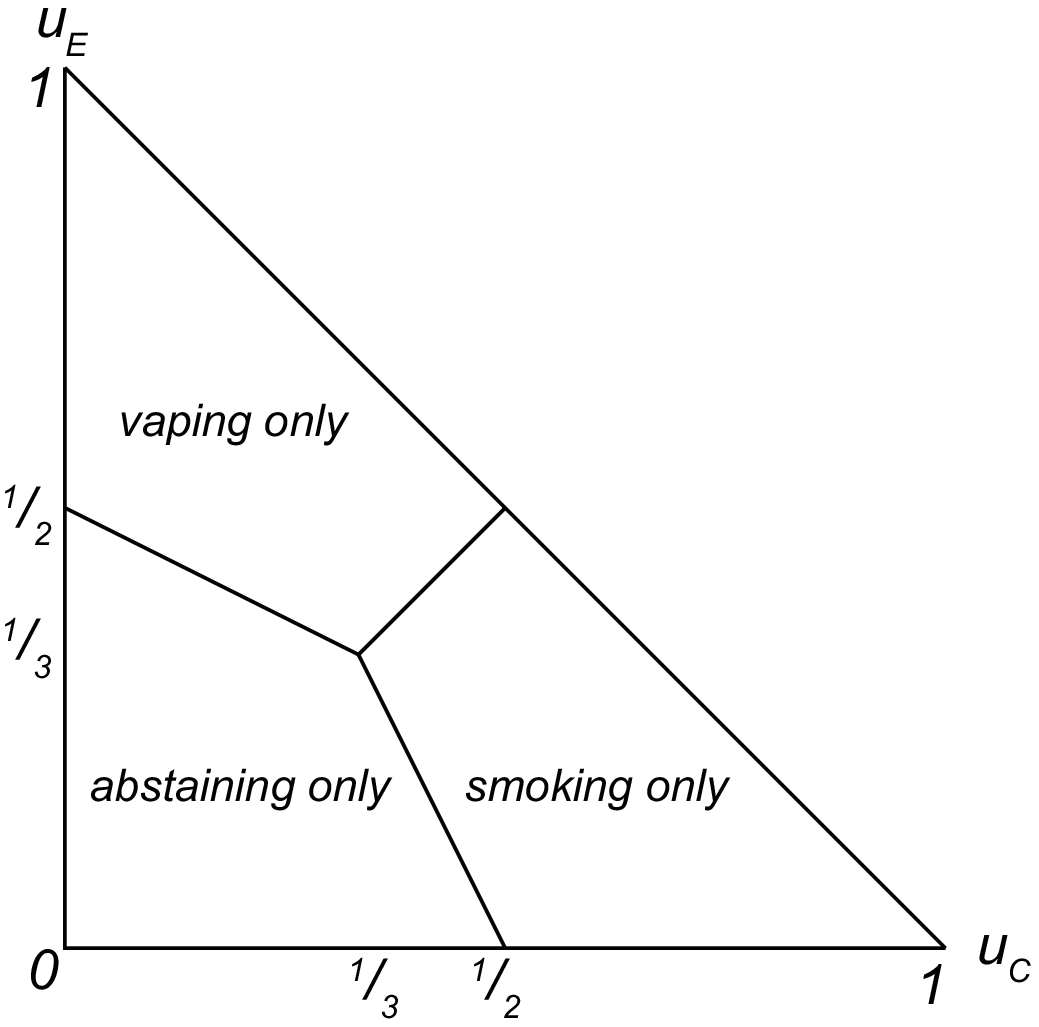}
    \caption{Stability regions for equilibria in the special case $\alpha=1$.}
    \label{fig:alpha1}
\end{figure}

When each equilibrium is substituted into the Jacobian, the eigenvalues of the matrix (see Table \ref{tab:alpha1}) give the stability. All eigenvalues are real, so if both eigenvalues are negative, the equilibrium is a stable node. The stability regions for the equilibria are shown in Figure \ref{fig:alpha1}.

\begin{table}[ht]
\caption{Equilibria and eigenvalues for system (\ref{eq:dCdt})-(\ref{eq:dEdt}) in the special case $\alpha=1$.}
\begin{tabular}{p{4cm} | p{4cm}}
Equilibrium & Eigenvalues \\ \hline
\begin{tabular}[c]{@{}l@{}}Abstaining only\\ $(C^*, E^*) = (0,0)$\end{tabular} & \begin{tabular}[c]{@{}l@{}}$\lambda_1 = 2u_C+u_E-1$\\ $\lambda_2 = u_C+2u_E-1$\end{tabular} \\ \hline
\begin{tabular}[c]{@{}l@{}}Smoking only\\ $(C^*, E^*) = (1,0)$\end{tabular}    & \begin{tabular}[c]{@{}l@{}}$\lambda_1 = u_E-u_C$\\ $\lambda_2 = 1-2u_C-u_E$\end{tabular}    \\ \hline
\begin{tabular}[c]{@{}l@{}}Vaping only\\ $(C^*, E^*) = (0,1)$\end{tabular}     & \begin{tabular}[c]{@{}l@{}}$\lambda_1 = u_C-u_E$\\ $\lambda_2 = 1-u_C-2u_E$\end{tabular}   
\end{tabular}
\label{tab:alpha1}
\end{table}

%%%%%%%%%%%%%%%%%%%%%%%%%%%%%%%%%%%%%%%%%%
%%%%%%%%%%%%%%%%%%%%%%%%%%%%%%%%%%%%%%%%%%

\subsection{Technical Details: Intrinsic Utility Function}
\begin{figure}[th]
    \centering
    \includegraphics[width=\textwidth]{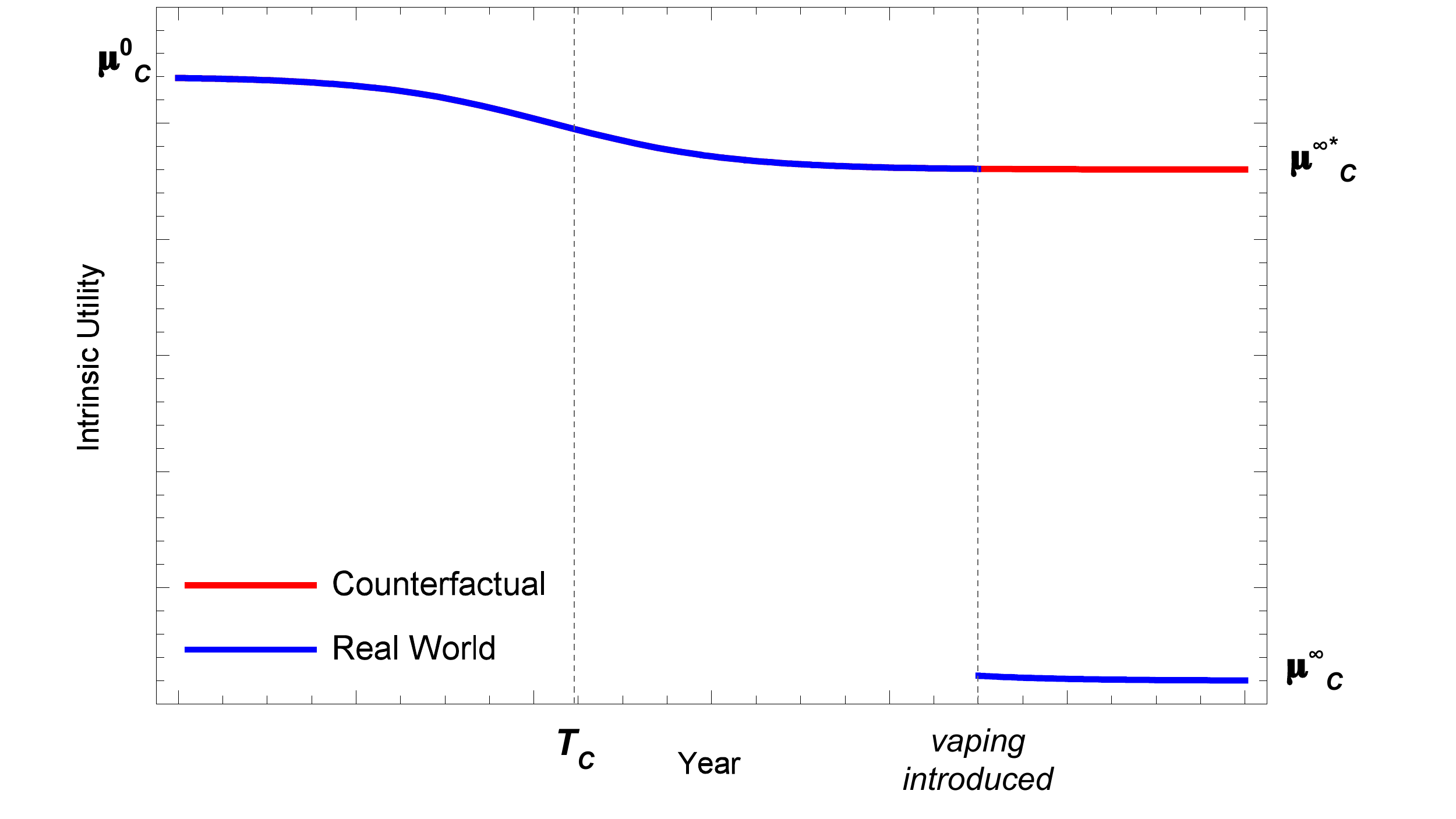}
    \caption{The sigmoidal function (Equation \ref{eq:ux}) describing the intrinsic utility of cigarettes for adults. As illustrated, the intrinsic utility of traditional smoking clearly decreases once vaping is introduced into society.}
    \label{fig:intrin}
\end{figure}

The intrinsic utility function (Equation \ref{eq:ux}) is sigmoidal. Following the introduction of e-cigarettes into society, the final intrinsic utility of traditional smoking changes (likely decreases) in response. Therefore, the intrinsic utility of cigarettes is a piecewise function (see Figure \ref{fig:intrin}):
\[  u_C(t) =
\begin{dcases}
u_C^{0}+\frac{u_C^{\infty*}-u_C^{0}}{1+ e^{-\lambda (t-T_C)}}, & t < \text{ year vaping introduced} \\
u_C^{0}+\frac{u_C^{\infty}-u_C^{0}}{1+ e^{-\lambda (t-T_C)}}, & t \ge \text{ year vaping introduced}
\end{dcases}
\]

%%%%%%%%%%%%%%%%%%%%%%%%%%%%%%%%%%%%%%%%%%
%%%%%%%%%%%%%%%%%%%%%%%%%%%%%%%%%%%%%%%%%%

\subsection{Technical Details: Sensitivity Analysis}
To test the sensitivity of model projections to the parameter $T_E$, we first set $T_E$ equal to each year in the plausible range 2022-2028 and hold all fixed parameter values constant. We then fit the model to the data using the algorithm specified in the main paper. 

\end{document}